\documentclass[useAMS,usenatbib]{mn2e}
\usepackage{graphicx}

\title[Elusive AGN]{Elusive Active Galactic Nuclei}
\author[Maiolino et al.]{R.~Maiolino$^{1}$, A.~Comastri$^{2}$,
R.~Gilli$^{1}$, N.~M.~Nagar$^{3}$, S.~Bianchi$^{4}$, T.~B\"{o}ker$^{5}$,
\newauthor
E.~Colbert$^{6}$,
A.~Krabbe$^{7}$, A.~Marconi$^{1}$, G.~Matt$^{4}$ and M.~Salvati$^{1}$
\\
$^{1}$INAF - Osservatorio Astrofisico di Arcetri, Largo E. Fermi 5, I-50125, Firenze, Italy\\
$^{2}$INAF - Osservatorio Astronomico di Bologna, via Ranzani 1, I-40126, Italy\\
$^{3}$Kapteyn Institute, University of Groningen,
        Landleven 12, 9747 AD Groningen, The Netherlands\\
$^{4}$Dip. di Fisica - Univ. degli Studi ``Roma III'', via della Vasca Navale 84,
        I-00146, Italy \\
$^{5}$Astrophysics Division, RSSD, European Space Agency, ESTEC, 
      NL-2200 AG Noordwijk, The Netherlands \\
$^{6}$Department of Physics and Astronomy, Johns Hopkins University,
        Baltimore, MD 21218, USA \\
$^{7}$University of California at Berkeley, 366 Le Conte Hall, Berkeley, CA 94720-7300, USA \\
}
\begin{document}
\date{Accepted . Received }
\pagerange{\pageref{firstpage}--\pageref{lastpage}} \pubyear{2002}
\maketitle
\label{firstpage}
\begin{abstract}
A fraction of active galactic nuclei do not show the classical
Seyfert-type signatures in their optical spectra, i.e. they are optically
``elusive''. X-ray observations are an optimal tool to identify this
class of objects. We combine new Chandra observations with archival
X-ray data in order to obtain a first estimate of the fraction
of elusive AGN in local galaxies and to constrain their nature.
Our results suggest that elusive AGN have a local density comparable 
to or even higher than optically classified Seyfert nuclei.
Most elusive AGN are heavily 
absorbed in the X-rays, with gas column densities exceeding 
$\rm 10^{24}cm^{-2}$, suggesting that their peculiar nature is 
associated with obscuration. It is likely that in elusive AGN,
the nuclear UV source is completely embedded and the ionizing 
photons cannot escape, which prevents the formation of a classical 
Narrow Line Region. Elusive AGN may contribute significantly to 
the 30~keV bump of the X-ray background.
\end{abstract}


\section{Introduction and definition of ``elusive AGN''}
There is growing evidence that the classification of galactic nuclei based
on their optical spectra alone provides an incomplete and sometimes 
deceiving description of their true nature. Indeed, some ``classical'' 
starburst galaxies show signatures of a hidden active galactic nucleus 
(AGN) at non-optical wavelengths. The most convincing evidence for 
obscured AGN hidden in starburst nuclei comes from X-ray observations. 
The prototype of this class of objects is NGC\,4945 which hosts a 
powerful nuclear starburst, identified through emission in the 
Br$\gamma$ and Pa$\alpha$ lines and through its mid-IR spectrum 
\citep{marconi00,moorwood96}. The observations 
indicate that a starburst superwind has created a large cavity where 
shocked clouds emit faint LINER-type lines. Studies at optical and 
near-IR wavelengths reveal no evidence for an AGN in this galaxy.
However, X-ray observations unambiguously prove the presence of a 
powerful and heavily obscured AGN: the 2-10 keV spectrum shows a
reflection-dominated component with a prominent Fe line, while the intrinsic
transmitted component is observed as a strong, variable excess in the
range 10-100 keV \citep{done96,guainazzi00}. Similar cases are NGC\,6240 
(Vignati et al. 1999), which has an optical spectrum very similar to 
that of NGC\,4945 \citep[i.e. weak LINER-type lines probably associated with 
the starburst superwind shocks;][]{lutz99}, 
and NGC\,3690 \citep{dellaceca02} which has an HII-type optical spectrum.
In all three galaxies, X-ray observations clearly indicate the presence
of an obscured active nucleus, yet their optical spectra
show no signs of the ``classical'' Seyfert-type emission lines.

We define ``optically elusive AGNs'', or simply ``elusive AGNs'',
those objects which do not show a Seyfert-like spectrum in the optical,
but which host a hard X-ray  nuclear source whose intrinsic
luminosity is in the Seyfert range, i.e. $\rm L_{2-10keV}>10^{41}erg~s^{-1}$.
Therefore, elusive AGN are generally hosted in nuclei optically classified either as
HII or LINER.

We emphasize that a LINER spectrum does not necessarily imply the presence of
an AGN with Seyfert-like luminosity. For instance, most
low-luminosity nuclei studied by \citet{ho01} with classical
LINER spectra are intrinsically
much weaker than classical Seyferts ($\rm L_{2-10keV}\approx 10^{38} erg~s^{-1}$).
Alternatively, LINER spectra may originate in starburst-driven shocks
\citep{heckman90,lutz99} and therefore be
totally unrelated with the presence of an AGN.
An object with a LINER-like spectrum but
$\rm L_{2-10keV}>10^{41}erg~s^{-1}$ will be therefore
considered as an elusive AGN.

We have started a program to search for elusive AGN in starburst galaxies
using infrared, radio and X-ray observations.
In this paper, we present the results of a small set of Chandra 
observations. In combination with literature X-ray data for other 
galaxies, we derive a first estimate of the fraction of elusive AGNs 
in the local universe and attempt to constrain their nature.

\section{The galaxy sample}
The parent sample consists of all non-Seyfert galaxies (typically starburst
with HII or LINER spectra) which show a high brightness temperature
($\rm T_b >10^5$K) in the VLBI observations obtained
by \citet{kewley00} and \citet{smith98a}.
The selection based on the presence
of a radio core was made to maximize the chance of finding an
elusive AGN.
Indeed, most Seyfert nuclei show a compact radio core with brightness temperature higher
than $\rm 10^5$K. However, brightness temperatures even in excess of $\rm 10^7$K
can also be achieved
by compact radio SNe \citep{smith98b}.
Therefore, the detection of a radio core alone is not a proof for the presence
of an AGN.

\citet{kewley00} selected nearby galaxies (z$<$0.025) from the IRAS catalog with
luminosity $\rm L_{IR}>10^{9.5}L_{\odot}$ (most of these
have $\rm L_{IR}<10^{11}L_{\odot}$),
and warm infrared colours: $\rm F_{60\mu m}/F_{100\mu m}>0.5$ {\it and}
$\rm F_{60\mu m}/F_{25\mu m}<8$. The warm colours are expected to favor
the selection of galaxies hosting AGN. The Kewley et al. sample contains 61
galaxies, 48 of which were observed with the Parkes Tidbinbilla Interferometer (PTI).
PTI detections
imply $\rm T_b>10^5$ K.

\citet{smith98a} selected the 40 most luminous
galaxies in the IRAS Bright Galaxy Sample ($\rm L_{IR}>10^{11.25}L_{\odot}$)
and observed with the VLBI
the subsample of the 31 galaxies with best VLA compact fluxes. 
Detections with the VLBI generally identify higher
brightness temperatures ($\rm T_b \gg 10^7$~K in several cases).

\section{X-ray identification of elusive AGN}
Since, by definition, elusive AGN cannot be identified
through the classical diagnostic diagrams involving optical
line ratios, we have to rely on X-ray--based classification
schemes. We infer the presence of an AGN if
one of the following conditions is met. i) for Compton-thin AGN:
an absorption-corrected X-ray {\it nuclear} luminosity of 
$\rm L_{2-10keV}>10^{41}erg~s^{-1}$ and a spectrum that can be 
described by an absorbed powerlaw with AGN-like slope ($\Gamma \sim 1.7$).
ii) for Compton-thick AGN:
a flat X-ray spectrum (powerlaw with $\Gamma < 1.0$) 
in the case of cold reflection\footnote{This limiting
value for $\Gamma < 1.0$ was chosen because most cold-reflection 
dominated, Compton-thick sources are in this range
\citep{bassani99, maiolino98} and also to discard
ULXs, which have $\Gamma > 1$ \citep{foschini02}.},
and/or an Fe line at 6.4-6.7~keV with equivalent width (EW) in
excess of 500 eV\footnote{This is the minimum value for a Compton-thick 
AGN found in \citet{bassani99}}, and/or the direct detection 
of the AGN emission at E$>$10keV (e.g. NGC\,4945). This definition is 
not as well-defined as the one adopted for the optical identification 
of Seyfert nuclei, and is subject to some ambiguities,
as discussed below.

\subsection{New Chandra observations}

We have used Chandra/ACIS-S to search for elusive AGN in four galaxies of 
our parent sample, namely UGC\,2369, NGC\,2623, NGC\,4418, and NGC\,4691. 
In addition, we include ACIS-S observations of NGC\,2993 which 
were serendipitously obtained during observations
of the companion NGC\,2992 (Colbert et al. in prep). The
data were reduced using CIAO v.2.3, following standard procedures.
Table~\ref{tab_chandra} gives a short observation log. 
In all cases, the nuclear X-ray source discussed here lies within 1$''$ 
of the position of the radio core or the near-IR peak, and thus can
confidently be identified with the galaxy nucleus. The X-ray fluxes from 
the nuclear sources are too weak to perform a detailed spectral 
analysis. For our analyses we used the C-statistics which is more
appropriate in the case of low signal-to-noise data \citep{cash79}.
 The weakest sources were fit with 
a simple power-law using a column density N$_H$ fixed to the 
Galactic value. For the brighter sources, the fit was improved by 
including a thermal component and/or by varying the value of intrinsic N$_H$. 
A summary of the best fit for each source is given in Table~\ref{tab_chandra}.

NGC\,2623 is the only source which unambiguously meets the above criteria for
the presence of an obscured AGN, since its very flat hard X-ray spectrum 
can only be explained by the presence of a Compton-thick, cold reflection 
dominated AGN (Fig.1).
Our data constrain the EW of the Fe line to an upper limit
of 1~keV. This value is somewhat lower than in typical Compton-thick AGN, 
but not unusual among other elusive AGN such as NGC\,4945 
(Bassani et al. 1999) or NGC\,3690 \citep{dellaceca02}. An absorbed
powerlaw with $\Gamma$ frozen to 1.9 gives a much worse fit.

NGC\,4418 also shows evidence for a flat spectrum emission component 
which may imply the presence of a Compton-thick AGN, but the limited
photon statistics make this identification somewhat tentative.
We note, however, that infrared observations also suggest
the presence of a heavily obscured AGN in this galaxy \citep{spoon01,dudley97}.
Also in this case, an absorbed
powerlaw with $\Gamma$ frozen to 1.9 gives a significantly worse fit.

   \begin{figure}
   \centering
   \includegraphics[angle=-90,width=6cm]{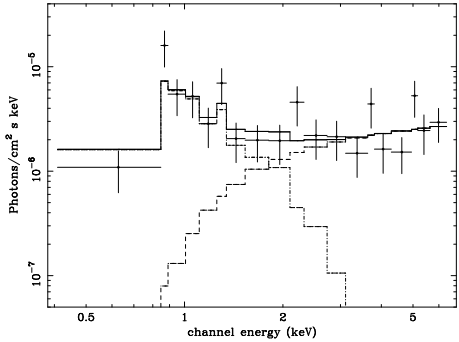}
   \caption{Unfolded Chandra spectrum of the nucleus of NGC\,2623 fitted
    with a powerlaw ($\rm \Gamma = -0.3$) and a thermal component
     ($\rm kT=0.6$~keV).}
              \label{n2623}%
    \end{figure}

   \begin{table}
      \caption[]{New Chandra observations of galaxies 
      with radio cores and optically not classified as Seyfert.}
     
         \begin{tabular}{lcccccc}
            \hline
            \noalign{\smallskip}
        Name      &  T$_{\rm int}^a$ &
	   cts$^b$ &  {F$_{\rm X}$}$^c$ & N$_{\rm H}^d$  & $\Gamma$ & kT\\
            \noalign{\smallskip}
            \hline
            \noalign{\smallskip}
             U2369 & 10 & 112 & 4.9 & $\rm 3^{+1}_{-1} $ & $\rm 1.3^{+0.9}_{-0.7}$ &
	              $\rm 0.8^{+0.1}_{-0.1}$\\
             N2623 & 20 & 129 & 13.0 & $\rm 9^{+4}_{-2}$ & $\rm -0.3^{+0.15}_{-0.16}$ &
	              $\rm 0.6^{+0.2}_{-0.1}$\\
             N2993 & 50 &  650 & 8.7 & $<$1.2 & $\rm 1.37^{+0.20}_{-0.20}$
	       &  -- \\
             N4418 & 20 & 25&  1.5 &  0.2$^e$ & $\rm 0.76^{+0.61}_{-0.62}$ & -- \\
             N4691 & 10 & 24&  1.4 &  0.2$^e$ & $\rm 1.91^{+0.63}_{-0.64}$ & -- \\
            \noalign{\smallskip}
            \hline
         \end{tabular}
$^a$ Integration time in ksec.\\
$^b$ Background subtracted counts.\\
$^c$ 0.3--8~keV observed flux in $\rm 10^{-14}~erg~s^{-1}cm^{-2}$.\\
$^d$ Absorbing column density in units of $\rm 10^{21}cm^{-2}$.\\
$^e$ Frozen to the Galactic value.
 \label{tab_chandra}
   \end{table}

\subsection{X-ray data from the literature}

We have collected additional hard X-ray data available
in the literature for other galaxies of our parent sample.
The objects for which hard X-ray data are available
are listed in Table~\ref{tab_tot}, where
we also report the optical classification of
the nuclei: HII and LINER are self-explanatory, while ``N4945''
indicates a spectrum similar to that of NGC\,4945, i.e. 
weak LINER-type lines around H$\alpha$, while H$\beta$ and [OIII]$\lambda$5007 
are undetected. In principle, the absence of H$\beta$ and [OIII]$\lambda$5007
does not allow one to distinguish between Sy2 and LINER. However, in NGC\,4945
\citet{moorwood96} showed that the overall spectral properties
are inconsistent with Sy2.  For the other
two objects with N4945-like spectra (namely NGC\,2623 and NGC\,4418)
the spectral coverage and quality of the optical spectra do not allow to
reach similarly firm conclusions. However, at least in the case of NGC\,4418,
for which \citet{lehnert95} present data with higher spectral resolution,
the width of the lines is more consistent with LINERs than Seyferts.
In the third column of Table~\ref{tab_tot},
we mark those objects that have X-ray spectra with evidence for 
an obscured AGN. In these cases, we
provide the absorption-corrected 2-10keV luminosity and the
column density of absorbing gas N$_H$. For Compton-thick AGN where  
only the reflection component is observed, we estimate
the intrinsic luminosity by assuming a reflection
efficiency of 1\% (with the exception of NGC\,253 discussed below).

Some additional explanation is required for NGC\,253 and NGC\,1808.
NGC\,253 was found to host a low-luminosity AGN with observed 
$\rm L_{2-10keV} = 10^{39}erg~s^{-1}$ which is obscured
by $\rm N_H \sim 2~10^{23}cm^{-2}$ \citep{weaver02}.
However, as discussed by Weaver et al., the detection of a 
broad radio recombination line by \citet{mohan02} strongly suggests
that the nucleus is absorbed by Compton-thick material and that
the intrinsic X-ray luminosity is $\sim$1000 times higher. 
NGC1808, on the other hand, appears to be variable:
in 1990, \citet{polletta96} found evidence for a mildly absorbed AGN
with $\rm L_{2-10keV}=2~10^{41}erg~s^{-1}$, but observations at later 
epochs \citep{bassani99} demonstrated that the nuclear source was
fainter by a factor of a few. It is possible that the elusive nature of 
this AGN is caused by fading rather than obscuration.

   \begin{table}
      \caption[]{Hard X-ray properties of galaxies with radio cores and optically
      not classified as Seyfert}
      
         \begin{tabular}{lcccc}
            \hline
            \noalign {\smallskip}
	    \multicolumn{5}{c}{$\rm 11 < lg (L_{IR}/L_{\odot}) < 12$}\\
         Name   &  Opt.$^a$ & X$^b$
	&  $\rm lg~L_X$$^c$ &  $\rm lg(N_H)$$^d$ \\
            \noalign{\smallskip}
            \hline
            \noalign{\smallskip}
	 UGC\,2369 &  HII$^{1,8}$    & ---$^i$ & &  \\
	 NGC\,2623 & N4945$^{1,8}$ & AGN$^i$ &
		$\approx$ 42.9$^e$ & $>$24 \\
	 UGC\,5101 & LINER$^{2,8,9}$ & AGN$^{ii}$ &
		$\approx$ 43.3$^e$ & $>$24 \\
	 NGC\,3690 &  HII$^{1,8}$   & AGN$^{iii}$ &
		 42.7 & 24.4  \\
     I1525+36 & LINER$^{2,8}$ & ---$^{iv}$ & &  \\
	 NGC\,6240 &  LINER$^{1,7}$ & AGN$^v$ &
		44.2   & 24.3 \\

            \noalign{\smallskip}
            \hline
            \noalign{\smallskip}

            \hline
            \noalign {\smallskip}
	    \multicolumn{5}{c}{$\rm lg (L_{IR}/L_{\odot}) < 11$}\\
         Name   &  Opt.$^a$ & X$^b$
	&   $\rm lg~L_X$$^c$ &  $\rm lg(N_H)$$^d$ \\
            \noalign{\smallskip}
            \hline
            \noalign{\smallskip}
	 NGC\,253  &  HII$^{7,12}$ & AGN(?)$^{vi}$ & 39~(42$^e$) & 23.3~($>$24) \\
	 NGC\,1672 & HII$^{3,7,10}$ & ---$^{vii}$ & &  \\
	 NGC\,1808 & HII$^{3,7,11}$ & AGN$^{viii}$ & 41.3 & 22.5 \\
	 NGC\,2993 & HII$^{4,7}$ & ---$^i$ & &   \\
	 NGC\,4418 & N4945$^{2,13}$ & AGN(?)$^i$ & (41.2$^e$) &  ($>$24) \\
	 NGC\,4691 & HII$^{7,5}$ & ---$^i$ & &   \\
            \noalign{\smallskip}
            \hline
            \noalign{\smallskip}
	 NGC\,4945 & N4945$^6$ & AGN$^{ix}$ & 42.8 & 24.6 \\
            \noalign{\smallskip}
            \hline

         \end{tabular}
$^a$ Optical classification (see text); references and notes:
$^1$\citet{wu98};
$^2$\citet{baan98};
$^3$\citet{vv86};
$^4$\citet{usui01};
$^5$\citet{garcia99};
$^6$\citet{moorwood96};
$^7$\citet{kewley00};
$^8$\citet{smith98a};
$^9$the integrated line ratios of UGC5101 are LINER-like ($^{2,8}$), but
\citet{concalves99} claim the detection of Sy-like optical signatures after a
multi-component fitting analysis;
$^{10}$the integrated line ratios of NGC1672 are HII-like ($^{7}$), but
\citet{vv81} claim the detection of Sy-like optical signatures after a
multi-component fitting analysis;
$^{11}$\citet{phillips93} dismissed previous claims for the presence of
Sy-like signatures in the optical spectrum;
$^{12}$\citet{moran96}; $^{13}$\citet{lehnert95}.\\
$^b$ X-ray evidence for an AGN is marked on this column.
References for the X-ray observations are:
$^i$ this work; $^{ii}$ \citet{ptak03};
 $^{iii}$ \citet{dellaceca02}; $^{iv}$ \citet{franceschini03};
 $^v$ \citet{vignati99}; $^{vi}$ \citet{weaver02};
 $^{vii}$ \citet{denaray00}; $^{viii}$ \citet{bassani99};
 $^{ix}$ \citet{guainazzi00}.\\
$^c$ Log of the absorption corrected 2--10~keV luminosity (erg~s$^{-1}$).\\
$^d$ Log of absorbing $\rm N_H$ in units of $\rm cm^{-2}$.\\
$^e$ Compton-thick objects for which the intrinsic X-ray luminosity
 is inferred by assuming a 1\% reflection efficiency (with the exception
 of NGC\,253, see text).
 \label{tab_tot}
   \end{table}

   \begin{figure}
   \centering
   \includegraphics[angle=0,width=8cm]{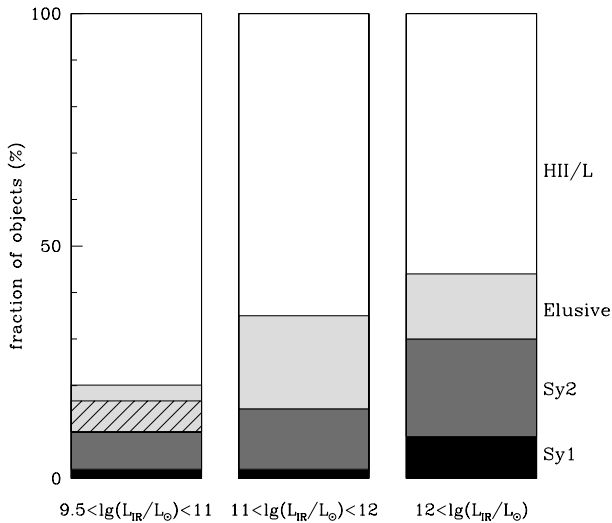}
   \caption{Estimated fraction of Seyfert and elusive AGN 
   in different infrared luminosity ranges. The hatched region in the
   lowest luminosity bin indicates the fraction of elusive AGN inferred from
   ambiguous cases.}
              \label{frac_el}%
    \end{figure}

\section{The fraction of elusive AGN}
Despite the limited statistics, we can make a first attempt 
at estimating the fraction of elusive AGN in the local universe.
As discussed by various authors \citep[e.g.][]{veilleux99, sanders96},
the fraction of AGN in galaxies increases with increasing infrared
luminosity.  Similarly to these studies, we divide our sample
by infrared luminosity and, more specifically,
into two bins:
$\rm 10^{9.5}<L_{IR}/L_{\odot}<10^{11}$ and $\rm 10^{11}< L_{IR}/L_{\odot}<10^{12}$.
For consistency with Veilleux et al., we adopt $\rm H_0=75~km~s^{-1}Mpc^{-1}$.

The $\bf 10^{9.5}<L_{IR}/L_{\odot}<10^{11}$ range. In this luminosity
range, the evidence for elusive AGN is more ambiguous than in
the higher luminosity bin. NGC\,4945 falls into this luminosity range: 
it is a clear case of an elusive AGN and is listed in Table~\ref{tab_tot} as a 
prototype for this class of objects. However, it is not included in our 
parent sample because its infrared colours are cooler than the selection 
threshold of \citet{kewley00}.
Therefore, NGC\,4945 will not be counted in the statistics used 
to derive the local fraction of elusive AGN.

 About 48\% of the local galaxies with $\rm 10^{9.5}<L_{IR}/L_{\odot}<10^{11}$
\citep[from the IRAS Bright Galaxy Survey,][]{sanders95}
have infrared colours matching the
selection criterion used by Kewley et al. (2000). We conservatively
assume that elusive AGN are only hosted in galaxies matching the Kewley et al.
IR colour criteria. In their sample,
42\% of the galaxies not classified as Seyfert host a radio core.
Of these, 6 galaxies have been observed in the hard X-rays (Table~\ref{tab_tot}) and 3
of them appear to host an elusive AGN, though 2 of these are dubious.
Therefore, a rough estimate of the fraction of elusive AGN in this
luminosity range is 48\%$\times$42\%$\times$3/6$\approx$10\%; if
we exclude the two dubious cases the fraction estimate drops to $\sim$3\%.
We summarize this result in Fig.~\ref{frac_el},
where we include the fraction of 
local galaxies with optically identified AGN derived by 
\citet{maiolino95}, i.e. 2\% for Sy1 and 8\% for Sy2\footnote{Most of
the Seyfert in the \citet{maiolino95} sample fall in this luminosity
range.
We do not use the fraction obtained by \citet{veilleux95} because their AGN
statistics in this luminosity range are too poor (two AGN).}.
While this calculation is affected by small-number statistics
and the uncertain observational evidence for elusive AGNs in this 
luminosity bin, we argue below that it should at least represent
a lower limit to the actual fraction of elusive AGN in the local
universe. 

 The $\bf 10^{11}< L_{IR}/L_{\odot}<10^{12}$ range. The
clearest cases for elusive AGN can be found in this luminosity range. 
Four out of the six objects observed in hard X-rays clearly host a heavily 
obscured AGN.
%
%

All galaxies in this luminosity range are from the \citet{smith98a}
sample, with the exception of NGC\,6240 and NGC1614 which are in
\citet{kewley00}. There are 34 galaxies in the parent sample {\it not} classified
as Seyfert, 25 of which were observed with VLBI.
 We conservatively assume that galaxies not
observed or not detected with VLBI do not contain elusive AGN.
Sixteen of the galaxies observed with VLBI
were found to have $\rm T_b > 10^5$K (about half of them have
$\rm T_b \gg 10^7$K).
Six of the galaxies with $\rm T_b > 10^5$K core were observed
in the hard X-rays (Table~\ref{tab_tot}) and 4 were found to host an obscured AGN.
Thus, the fraction of X-ray detected elusive AGN in this luminosity range can be estimated 
as $16/34 \times 4/6 = 0.31$.
This is the estimated fraction of elusive AGN among galaxies
{\it not} classified as Seyfert. \citet{veilleux95} estimate that 
the fraction of optical HII/LINER (i.e. {\it non}-Seyfert) galaxies in
this luminosity range is 85\%. Therefore the estimated fraction of elusive AGN
in this luminosity range should be 31\%~$\times$~0.85~$\approx$ 26\% of all galaxies. 
However, we have to apply another correction to account for the
lack of galaxies with luminosity $\rm <10^{11.25}$ in the Smith et al. 
sample. By interpolating with the lower luminosity bin discussed above, we 
finally estimate that the total fraction of elusive AGN in the luminosity 
range $\rm 10^{11}< L_{IR}/L_{\odot}<10^{12}$ is $\approx$20\%.
We again illustrate this result in Fig.~\ref{frac_el} together with the fraction 
of Seyfert galaxies identified optically by \citet{veilleux99}, i.e. 
2\% for Sy1 and 13\% for Sy2.

{\bf ULIRGs}. 
In this paper we do not discuss Ultraluminous Infrared Galaxies (ULIRGs,
$\rm L_{IR}/L_{\odot}>10^{12}$), since they are much sparser and poorly
representative of the local population of galaxies. However, 
for completeness, we also include in Fig.~\ref{frac_el} the fraction of 
elusive AGN among ULIRGs as derived from X-ray observations in the 
literature \citep{franceschini03, ptak03, 
risaliti00}\footnote{Note that the IR luminosities
of the galaxies were adjusted to $\rm H_0=75~km~s^{-1}Mpc^{-1}$,
the value adopted for this paper.}
along with the fraction of Seyferts optically identified by \citet{veilleux99}.
Combining these data we obtain:
Sy1=9\%, Sy2=21\% and elusive AGN=14\%.

We note that the fractions of elusive AGN inferred above are probably 
lower limits because we have assumed that galaxies
not matching the selection criteria of Kewley et al. and Smith et al. 
do {\it not} host elusive AGN. The obvious counter example of 
NGC\,4945 demonstrates that this assumption is a conservative one. 
Moreover, there are some examples of classical Seyfert nuclei without a
detected radio core,
and therefore elusive AGN without a radio core may also exist.
Finally, a large fraction of the elusive AGN in Table~\ref{tab_tot} are relatively 
well-studied objects with better X-ray data (longer integrations and 
more detailed analysis). Elusive AGN may have been missed in other 
objects which have received less attention.


\section{The nature of elusive AGN}
It is not clear why elusive AGNs
do not produce the typical Seyfert-type emission lines in their optical 
spectra.
Dilution by the circumnuclear region or by the host galaxy
may be an explanation; in particular \citet{moran02} have shown that
this may be the case for some of the X-ray selected AGN at high redshift.
However, most of the
objects in our sample are very nearby and have been studied in great detail, and
generally no optical Seyfert-like signatures were found (but see
notes to Table~\ref{tab_tot}).
A lack of an ``UV bump'' typically observed (or inferred)
in Seyferts is also unlikely: at least some elusive AGN show 
strong emission from hot dust (\cite{genzel98} for UGC\,5101,
\cite{krabbe01} for NGC\,4945, \cite{maiolino03} for NGC\,3690)
which must be heated by the UV bump associated with the X-ray emission.

\begin{figure}
   \centering
   \includegraphics[angle=0,width=7cm]{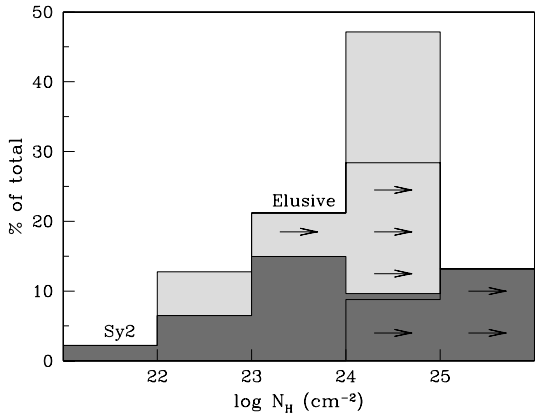}
   \caption{Cumulative distribution of absorbing N$_H$ for Sy2's and
   elusive AGN's, under the assumption that elusive AGN's are as numerous
   as Sy2's. The dark-gray histogram shows the contribution by Sy2's, while the
   light-gray histogram shows the contribution by elusive AGN's. Histograms
   marked with arrows show the fraction of objects with a lower limit on N$_H$}
              \label{Nh}%
    \end{figure}

The X-ray spectra of elusive AGN provide some hints of their nature. 
Indeed, most elusive AGN are Compton-thick ($\rm N_H >10^{24}cm^{-2}$).
Fig.~\ref{Nh} shows the cumulative N$_H$ distribution of elusive
AGN and Sy2 \citep[the latter from][]{risaliti99}\footnote{We removed NGC4945 and NGC1808
from the sample of Risaliti et al., which were included because of their
Seyfert-like X-ray spectra and not because of their optical classification.},
under the assumption that elusive AGN and Sy2 have the same local density.
The comparison of the N$_H$
distributions indicates that elusive AGN are more absorbed than Sy2.
This strongly suggests that the elusive nature of this
class of nuclei is associated with heavy obscuration.

One possibility
is that the Narrow Line Region (NLR) is also heavily obscured along
our line of sight. However, spectroscopic observations in the near-IR
and mid-IR, where dust extinction is much reduced, did not detect
the expected narrow emission lines in several of these
objects \citep[Maiolino et al. in prep.]{marconi94, spoon00, genzel98}. Moreover, the
NLR is generally too extended to be completely obscured.

Another possibility is that the nuclear pc-scale absorber is not distributed
in the torus-like geometry typically assumed for Seyfert~2 nuclei,
but covers the nuclear UV source in all directions. 
In this case, the UV photons cannot escape to produce
a classical NLR. Detailed studies of NGC\,4945
favor this scenario \citep{marconi00,moorwood96}.
Indeed, the
clouds located in the cavity produced by the nuclear starburst
superwind are very faint and only weakly ionized, implying that they are not
exposed to the strong photoionizing UV continuum which is expected to be
associated with the X-ray emission. As suggested by Done et al. (2003)
a disk/torus of Compton-thick material may extend (though
with lower $\rm N_H$) to high latitudes and totally cover the AGN.
A similar scenario was also proposed by \citet{dudley97},
who predict that such embedded AGN should be characterized by
a deep Silicate absorption at 9.7$\mu$m.  Notably, our sample includes
some of the objects with the deepest Silicate absorption, e.g. NGC4945
\citep{maiolino00} and NGC4418 \citep{spoon01}.


\section{Conclusions and discussion}
By combining new and past hard X-ray observations we have defined
a small sample of optically elusive AGN, i.e. nuclei with optical
spectra that do not show any evidence for the presence of
a Seyfert nucleus, but with X-ray properties indicative of a 
heavily obscured AGN (with Seyfert-type luminosity). After accounting for
selection effects, we estimate that the fraction of galaxies with
elusive AGN may be relatively high, i.e. comparable to or higher then
classical, optically classified Seyfert nuclei.
As a consequence,
the local total density of AGN may be a factor of $\sim$2 higher than 
estimated from optical spectroscopic surveys.
Our results also imply that the ratio between obscured and unobscured AGN
is higher than estimated previously, i.e. about 7:1 in the luminosity range
$\rm log(L_{IR}/L_{\odot})<11$ and about 16:1 in the luminosity range
$\rm 11<log(L_{IR}/L_{\odot})<12$.

Nearly all elusive AGN are heavily absorbed in the X-rays (Compton thick),
suggesting that their elusive nature is associated with heavy obscuration.
The geometry of the obscuring medium may be different than in Sy2 galaxies. 
In particular, the extension of the Compton thick medium may completely
embed the nuclear source, preventing
UV photons from escaping and producing the Narrow Line Region.

The X-ray spectral properties of elusive AGN may have important implications
for the X-ray background. Indeed, while weak at E$<$10~keV (because
of the strong absorption), the X-ray spectral energy distribution of those
elusive AGN observed above 10~keV peaks at about 30~keV,
due to an absorbing column
density in the range $\rm 10^{24}<N_H<10^{25}cm^{-2}$, and therefore
may provide an important contribution to the peak of the
X-ray background at 30 keV.
Most of these elusive AGN would 
not be detected below 10~keV in Chandra and XMM surveys, except in
the X-ray brightest cases. Such sources may be the so-called X-ray
Bright Optically Normal Galaxies \citep[XBONGs,]{comastri02}.
Yet, an assessment of the contribution of elusive AGN to the
30~keV background bump requires a better knowledge of their luminosity
distribution.  An XMM program aimed at studying a sample
of candidate local elusive AGN, recently approved for AOT3, will help
to tackle these issues. 

\section*{Acknowledgements}
We are grateful to D. Alexander, M. Veron and P. Veron for useful comments.
This work was partially supported by the Italian Space Agency (ASI) under grants
I/R/073/02 and I/R/057/02 and by the
Italian Ministry for University and Research (MIUR) under grant Cofin-00-2-35.

\label{lastpage}

\begin{thebibliography}{}

\bibitem[\protect\citeauthoryear{Baan et al.}{1998}]{baan98} Baan, W., Salzer, J.,
Lewinter, R.,\ 1998, ApJ, 509, 633

\bibitem[\protect\citeauthoryear{Bassani et al.}{1999}]{bassani99} Bassani, L., et al.\
1999, ApJS, 121, 473 


\bibitem[Cash(1979)]{cash79} Cash, W.\ 1979, ApJ, 228, 939 


\bibitem[\protect\citeauthoryear{Comastri et al.}{2002}]{comastri02} Comastri, A., et al.
in ``New Visions of the X-ray Universe'', in press (astro-ph/0203019)

\bibitem[Gon{\c c}alves, V{\' e}ron-Cetty, \& V{\' 
e}ron(1999)]{concalves99} Gon{\c c}alves, A.~C., V{\' e}ron-Cetty, 
M.-P., \& V{\' e}ron, P.\ 1999, A\&AS, 135, 437 

\bibitem[\protect\citeauthoryear{Della Ceca et al.}{2002}]{dellaceca02} Della Ceca, R.~et 
al.\ 2002, ApJ, 581, L9 

\bibitem[\protect\citeauthoryear{de Naray et al.}{2000}]{denaray00} 
de Naray, P.~J., Brandt, W.~N., Halpern, J.~P., \& Iwasawa, K.\ 2000, AJ, 
119, 612

\bibitem[\protect\citeauthoryear{Done et al.}{1996}]{done96} Done, C., 
Madejski, G.~M., \& Smith, D.~A.\ 1996, ApJ, 463, L63 

\bibitem[\protect\citeauthoryear{Done et al.}{2003}]{done03} Done, C., Madejski, G.~M., Zycki, P.~T., \& Greenhill, L.~J.\ 2003, ApJ, 588, 763 

\bibitem[\protect\citeauthoryear{Dudley \& Wynn-Williams}{1997}]{dudley97} Dudley, 
C.~\& Wynn-Williams, C.\ 1997, ApJ, 488, 720 



\bibitem[\protect\citeauthoryear{Foschini et al.}{2002}]{foschini02}Foschini, L.~et al.\ 
2002, A\&A, 392, 817  

\bibitem[\protect\citeauthoryear{Franceschini et al.}{2003}]{franceschini03} Franceschini, A.,
et al. 2003, MNRAS (astro-ph/0304529)

\bibitem[\protect\citeauthoryear{Garc{\'{\i}}a-Barreto et al.}{1999}]{garcia99} 
Garc{\'{\i}}a-Barreto, J.~A., Aceves, H., Kuhn, O., Canalizo, G., Carrillo, 
R., \& Franco, J.\ 1999, Revista Mexicana de Astronomia y Astrofisica, 35, 
173

\bibitem[\protect\citeauthoryear{Genzel et al.}{1998}]{genzel98} Genzel, R.~et al.\ 1998, 
ApJ, 498, 579 

\bibitem[\protect\citeauthoryear{Guainazzi et al.}{2000}]{guainazzi00} Guainazzi, M., et al.\
2000, A\&A, 356, 463





\bibitem[Heckman, Armus, \& Miley(1990)]{heckman90} Heckman, 
T.~M., Armus, L., \& Miley, G.~K.\ 1990, ApJS, 74, 833

\bibitem[\protect\citeauthoryear{Ho et al.}{2001}]{ho01} Ho, L.~C.~et al.\ 2001, 
ApJ, 549, L51 



\bibitem[\protect\citeauthoryear{Kewley et al.}{2000}]{kewley00} Kewley, L.~J., et al.\
2000, ApJ, 530, 704 

\bibitem[\protect\citeauthoryear{Krabbe et al.}{2001}]{krabbe01} 
Krabbe, A., B{\" o}ker, T., \& Maiolino, R.\ 2001, ApJ, 557, 626 



\bibitem[Lehnert \& Heckman(1995)]{lehnert95} Lehnert, M.~D.~\& 
Heckman, T.~M.\ 1995, ApJS, 97, 89 

\bibitem[\protect\citeauthoryear{Lutz et al.}{1999}]{lutz99} Lutz, D., 
Veilleux, S., \& Genzel, R.\ 1999, ApJ, 517, L13 

\bibitem[\protect\citeauthoryear{Maiolino \& Rieke}{1995}]{maiolino95} Maiolino, R.~\& 
Rieke, G.~H.\ 1995, ApJ, 454, 95

\bibitem[\protect\citeauthoryear{Maiolino et al.}{1998}]{maiolino98}Maiolino, R., Salvati, 
M., Bassani, L., et al.\ 1998, A\&A, 338, 781 

\bibitem[\protect\citeauthoryear{Maiolino et al.}{2000}]{maiolino00} Maiolino, R., et al.\
 2000, ASP Conf.~Ser.~195: 
Imaging the Universe in Three Dimensions, 307 

\bibitem[\protect\citeauthoryear{Maiolino}{2003}]{maiolino03} Maiolino, R., 
 2003, ASP Conf.~Ser.~290: 
Active Galactic Nuclei: From Central Engine to Host Galaxy, 457 

\bibitem[\protect\citeauthoryear{Marconi et al.}{2000}]{marconi00} Marconi, A., et al.\
2000, A\&A, 357, 24

\bibitem[\protect\citeauthoryear{Marconi et al.}{1994}]{marconi94} 
Marconi, A., Moorwood, A.~F.~M., Salvati, M., \& Oliva, E.\ 1994, A\&A, 
291, 18

\bibitem[\protect\citeauthoryear{Mohan et al.}{2002}]{mohan02} Mohan, 
N.~R., Anantharamaiah, K.~R., \& Goss, W.~M.\ 2002, ApJ, 574, 701

\bibitem[Moran, Filippenko, \& Chornock(2002)]{moran02} Moran, 
E.~C., Filippenko, A.~V., \& Chornock, R.\ 2002, ApJL, 579, L71

\bibitem[\protect\citeauthoryear{Moorwood et al.}{1996}]{moorwood96} Moorwood, A.~F.~M., 
van der Werf, P.~P., Kotilainen, J.~K., Marconi, A., \& Oliva, E.\ 1996, 
A\&A, 308, L1

\bibitem[Moran, Halpern, \& Helfand(1996)]{moran96} Moran, 
E.~C., Halpern, J.~P., \& Helfand, D.~J.\ 1996, ApJS, 106, 341

\bibitem[Phillips(1993)]{phillips93} Phillips, A.~C.\ 1993, AJ, 
105, 486

\bibitem[\protect\citeauthoryear{Polletta et al.}{1996}]{polletta96} Polletta, M., Bassani, 
L., Malaguti, G., Palumbo, G.~G.~C., \& Caroli, E.\ 1996, ApJS, 106, 399 

\bibitem[\protect\citeauthoryear{Ptak et al.}{2003}]{ptak03} Ptak, A., Heckman, T., Levenson, N.~A.,
Weaver, K., Strickland, D., 2003, ApJ, in press (astro-ph/0304222)

\bibitem[\protect\citeauthoryear{Risaliti et al.}{1999}]{risaliti99} Risaliti, G., Maiolino, R., \& Salvati, M.\ 1999, ApJ, 522, 157

\bibitem[\protect\citeauthoryear{Risaliti et al.}{2000}]{risaliti00} 
Risaliti, G., Gilli, R., Maiolino, R., \& Salvati, M.\ 2000, A\&A, 357, 13

\bibitem[Sanders et al.(1995)]{sanders95} Sanders, D.~B., Egami, 
E., Lipari, S., Mirabel, I.~F., \& Soifer, B.~T.\ 1995, AJ, 110, 1993 


\bibitem[\protect\citeauthoryear{Sanders \& Mirabel}{1996}]{sanders96} Sanders, D.~B.~\& 
Mirabel, I.~F.\ 1996, ARA\&A, 34, 749 

\bibitem[\protect\citeauthoryear{Smith et al.}{1998a}]{smith98a} Smith, 
H.~E., Lonsdale, C.~J., \& Lonsdale, C.~J.\ 1998, ApJ, 492, 137 

\bibitem[\protect\citeauthoryear{Smith et al.}{1998b}]{smith98b} Smith, H.~E., Lonsdale, C.~J., Lonsdale, C.~J., \& Diamond, P.~J.\ 1998, ApJ, 493, L17 



\bibitem[\protect\citeauthoryear{Spoon et al.}{2001}]{spoon01} Spoon, H.~W.~W., Keane, 
J.~V., Tielens, A.~G.~G.~M., Lutz, D., \& Moorwood, A.~F.~M.\ 2001, A\&A, 
365, L353 

\bibitem[\protect\citeauthoryear{Spoon et al.}{2000}]{spoon00} Spoon, H.~W.~W., 
Koornneef, J., Moorwood, A.~F.~M., Lutz, D., \& Tielens, A.~G.~G.~M.\ 2000, 
A\&A, 357, 898

\bibitem[Usui, Sait{\= o}, \& Tomita(2001)]{usui01} Usui, T., 
Sait{\= o}, M., \& Tomita, A.\ 2001, AJ, 121, 2483

\bibitem[\protect\citeauthoryear{Veilleux et al.}{1995}]{veilleux95} Veilleux, S., et al.\
1995, ApJS, 98, 171

\bibitem[\protect\citeauthoryear{Veilleux et al.}{1999}]{veilleux99} Veilleux, 
S., Kim, D., \& Sanders, D.\ 1999, ApJ, 522, 113 

\bibitem[Veron, Veron, \& Zuiderwijk(1981)]{vv81} Veron, 
M.~P., Veron, P., \& Zuiderwijk, E.~J.\ 1981, A\&A, 98, 34

\bibitem[Veron-Cetty \& Veron(1986)]{vv86} Veron-Cetty, 
M.-P.~\& Veron, P.\ 1986, A\&AS, 66, 335 

\bibitem[\protect\citeauthoryear{Vignati et al.}{1999}]{vignati99} Vignati, P.~et al.\ 
1999, A\&A, 349, L57 

\bibitem[\protect\citeauthoryear{Weaver et al.}{2002}]{weaver02} 
Weaver, K.~A., Heckman, T.~M., Strickland, D.~K., \& Dahlem, M.\ 2002, 
ApJ, 576, L19 

\bibitem[\protect\citeauthoryear{Wu et al.}{1998}]{wu98} Wu, H., Zou, 
Z.~L., Xia, X.~Y., \& Deng, Z.~G.\ 1998, A\&AS, 127, 521


\end{thebibliography}
\end{document}